\documentclass[aps, prl, twocolumn, superscriptaddress, preprintnumbers]{revtex4-1}

\usepackage{amsmath}	
\usepackage{amsthm}		
\usepackage{amssymb}	
\usepackage{eufrak}
\usepackage{datetime}
\usepackage{graphicx}
\usepackage{color}
\usepackage{verbatim}
\usepackage{bm}
\usepackage{float}

\usepackage[colorlinks=true, citecolor=midblue, linkcolor=midblue, urlcolor=midblue]{hyperref}

\usepackage{setspace}

\settimeformat{ampmtime}

\definecolor{grey}{rgb}{0.4,0.4,0.4}
\definecolor{dullmagenta}{rgb}{0.4,0,0.4}
\definecolor{darkblue}{rgb}{0,0,0.4}
\definecolor{midblue}{rgb}{0,0,0.5}
\definecolor{midred}{rgb}{0.5,0,0}
\definecolor{orange}{rgb}{1,0.5,0}
\definecolor{lightbrown}{rgb}{0.75,0.5,0.25}
\definecolor{tan}{cmyk}{0.14,0.42,0.56,0}
\definecolor{djunglegreen}{cmyk}{0.99,0,0.52,0}
\definecolor{lightgreen}{rgb}{0,1,0}
\definecolor{olivegreen}{cmyk}{0.64,0,0.95,0.40}
\definecolor{midgreen}{rgb}{0.0,0.675,0.0}
\definecolor{darkgreen}{rgb}{0,0.5,0}

\newcommand{\q}{\quad}
\newcommand{\qq}{\qquad}

\newcommand{\vs}{\vspace}
\newcommand{\hs}{\hspace}
\renewcommand{\.}{\hspace{0.5mm}}

\newcommand{\ra}{\ensuremath{\rightarrow}}

\newcommand{\Erm}{\ensuremath{\mathrm{E}}}

\newcommand{\Hrm}{\ensuremath{\mathrm{H}}}

\newcommand{\crm}{\ensuremath{\mathrm{c}}}
\newcommand{\drm}{\ensuremath{\mathrm{d}}}
\newcommand{\erm}{\ensuremath{\mathrm{e}}}

\newcommand{\grm}{\ensuremath{\mathrm{g}}}

\newcommand{\srm}{\ensuremath{\mathrm{s}}}

\newcommand{\Ncal}{\ensuremath{\mathcal{N}}}
\newcommand{\Ocal}{\ensuremath{\mathcal{O}}}

\newcommand{\Nbb}{\ensuremath{\mathbb{N}}}

\newcommand{\Rbb}{\ensuremath{\mathbb{R}}}

\renewcommand{\d}{\ensuremath{\mathrm{d}}}

\newcommand{\eg}{e.g.}
\newcommand{\ie}{i.e.}

\newcommand{\cf}{c.f.}

\def\Msun{M_\odot}

\settimeformat{ampmtime}

\usepackage{float}

\begin{document}

\title{Extreme-Value Distributions and Primordial Black-Hole Formation}

\author{Florian K{\"u}hnel}
\email{Florian.Kuehnel@physik.uni-muenchen.de}
\affiliation{
	Arnold Sommerfeld Center,
	Ludwig-Maximilians-Universit{\"a}t,
	Theresienstra{\ss}e 37,
	80333 M{\"u}nchen,
	Germany}

\author{Dominik J.~Schwarz}
\email{dschwarz@physik.uni-bielefeld.de}
\affiliation{
	Fakult{\"a}t f{\"u}r Physik,
	Universit{\"a}t Bielefeld,
	Postfach 100131,
	33501 Bielefeld,
	Germany}

\date{\formatdate{\day}{\month}{\year}, \currenttime}

\begin{abstract}
\vs{-1mm}
\hs{-2mm}We argue that primordial black-hole formation must be described by means of extreme-value theory. This is a consequence of the large values of the energy density required to initiate the collapse of black holes in the early Universe and the finite duration of their collapse. Compared to the Gau{\ss}ian description of the most extreme primordial density fluctuations, the holes' mass function is narrower and peaks towards larger masses. Secondly, thanks to the shallower fall-off of extreme-value distributions, the predicted abundance of primordial black holes is boosted by $10^{7}$ orders of magnitude when extrapolating the observed nearly scale-free power spectrum of the cosmic large-scale structure to primordial black-hole mass scales. 
\end{abstract}


\maketitle

{\it Introduction\;---\;}Primordial black holes (PBHs) \cite{ZeldovichNovikov69, Carr:1974nx} are black holes produced in the very early Universe, before the time of matter-radiation equality. The r{\^o}le of PBHs as dark-matter candidates has long been discussed \cite{1975Natur.253..251C}, see Ref.~\cite{Carr:2020xqk} for a recent review. After the detection of gravitational waves from coalescing black hole binaries \cite{TheLIGOScientific:2016pea}, it was quickly realised that these holes could conceivably be primordial in nature \cite{Bird:2016dcv}. 

PBHs can form when Hubble horizon sized overdensities collapse. This happens when the energy density contrast $\delta$ is above a (medium- and shape-dependent) critical threshold $\delta_{\crm}$. During the radiation-dominated epoch, and assuming sphericity of the overdensity, a value of $\delta_{\crm} = 0.45$ has been obtained using numerical simulations \cite{Musco:2012au}, \footnote{The numerical value of the threshold $\delta_{\crm}$ depends on shape and statistics of the collapsing overdensities, \cf~Refs. \cite{Escriva:2019phb, Musco:2020jjb} for spherical perturbations and Ref.~\cite{Kuhnel:2016exn} for non-spherical shapes. Furthermore, there is a slight discrepancy amongst the results of various groups.}. However, it has been demonstrated that this value changes throughout the cosmic history, most dominantly during the QCD transition \cite{Crawford:1982yz}, but also essentially under all circumstances of a reduction of the sound speed \cite{Carr:2019kxo}, yielding enhanced PBH formation at the corresponding horizon mass scales.

As the value of the critical overdensity is huge as compared to that of typically measured fluctuations in the cosmic microwave background, $\delta_{\rm rms} = 10^{-4}$ at the horizon scale at photon decoupling \cite{Akrami:2018odb}, the amount of produced PBHs crucially depends on the tail behaviour of the density distribution. Therefore, PBH formation is an inherently rare event. For instance, if all the dark matter was constituted by PBHs with mass of $10^{20}\,\grm$, which would be formed when the Universe had a temperature of $T \sim 10^{4}$\,TeV, only one out of $10^{15}$ horizon patches at that temperature would have to match the condition for yielding black-hole formation. Once the collapse is initiated, it is far from being instantaneous{\;---\;}usually taking a number of e-folds till the hole has finally formed \cite{Musco:2008hv, Musco:2012au}.

The purpose of the {\it Letter} is to properly account for both of these characteristics of PBH formation{\;---\;}its rareness and its finite formation time. We will see that the relevant statistical distributions are far from being Gau{\ss}ian, even if this was true for the random variables describing the initial density contrast.

Modern cosmology is successfully based on the cosmological principle and has been tested by a large variety of cosmological observations, e.g.~\cite{Akrami:2018odb}. An immediate consequence of that principle is that density fluctuations on horizon sized patches can be described as independent and identically distributed (iid) random variables. In the following we focus on statistical arguments. For the sake of simplicity, we shall assume that the collapse of overdensities to PBHs is initiated at the same cosmic time throughout the Universe.

{\it Averages versus maxima\;---\;}We describe the fluctuations of the density contrast within a Hubble patch by a set of random variables $\{ \Delta_{i} \}_{i\mspace{2mu}=\mspace{2mu}1}^{K}$, where $K \in \Nbb$ is the total number of Hubble patches in the early Universe which constitute our present visible Universe. Furthermore, we decompose the total number of Hubble patches into regions of $N$ patches, with $1 \ll N \ll K$, such that each region contains a very large number of patches, but has (up to negligible corrections) at most one patch in which the threshold for PBH formation is exceeded. Those regions or blocks{\;---\;}their number being $B \equiv K / N${\;---\;}naturally correspond to a Hubble volume at a time later than the time at which the PBH forming overdensity re-enters the horizon.

Certainly, the probability that within each block no black hole forms is equal to the probability that the block maximum is below the black-hole formation threshold:
\vs{1mm}
\begin{subequations}
\begin{align}
	{\rm Prob}( \text{\it no$\mspace{-3mu}$ BH}\. )
		&=
					{\rm Prob}
					\bigg(
						\underset{i\.=\.1,\.\ldots,\.N}
							{\max\!
							\big(
								\Delta_{i}
							\big)}
						<
						\delta_{\crm}
					\bigg)
					\, .
					\label{eq:noPBH}
\intertext{Then,\vs{-2.5mm}}
		\tilde{\beta}_{N}
			&\equiv
					1
					-
					{\rm Prob}( \text{\it no$\mspace{-3mu}$ BH}\. )
					\label{eq:AtleastonePBH}
\end{align}
\end{subequations}
is the probability that (at least) one black hole forms. Note that if this quantity is much less than unity, it is related to the fraction of horizon-sized regions collapsing to PBHs, being commonly denoted by $\beta$ \cite{Carr:1975qj}, via
\begin{align}
	\beta
		&\simeq
					\tilde{\beta}_{N} / N
					\; .
					\label{eq:beta=betatilde/N}
\end{align}

Hence, for PBH formation{\;---\;}contrary of using statistics for {\it averages}{\;---\;}one needs to use statistics of {\it extremes}, \ie, of maxima in our case. Such questions are addressed by the well-established field of extreme-value theory, which dates back to the 1920's \cite{fisher_tippett_1928} (see Ref.~\cite{10.5555/262578} for a classic text book), and which is commonly used in numerous fields of risk management{\;---\;}essentially in any case the statistics of extremely-rare events matter. PBH formation precisely fits into this category, and so in this {\it Letter} we will apply extreme-value statistics to study the distribution of PBH.

In order to underline the difference amongst the two mentioned statistical problems, let us recapitulate the central limit theorem. This implies that the probability distribution of the {\it sample average},
\begin{align}
	S_{N}
		&\equiv
					\frac{ 1 }{ N } \sum_{i\mspace{2mu}=\mspace{2mu}1}^{N} \Delta_{i}
					\; ,
					\label{eq:sample-average}
\end{align}
of iid random variables $\{ \Delta_{i} \}_{i\mspace{2mu}=\mspace{2mu}1}^{N}$ converges to the Gau{\ss} or Normal distribution $\Ncal( 0, 1 )$ in the following sense:
\begin{align}
	\underset{N\mspace{2mu}\ra\mspace{2mu}\infty}
		{\lim}\.
		{\rm Prob}\!
		\left(
			\frac{ 
				S_{N}
				-
				\mu }
			{ \sqrt{\sigma^{2} / N\,} }
			<
			\delta
		\right)
		&=
					\Phi( \delta )
					,
					\label{eq:Normal-CDF}
\end{align}
with mean $\mu$ and variance $\sigma$, and the cumulative distribution function (CDF) of the Normal distribution is $\Phi( \delta ) \equiv 1 / \sqrt{2\pi\.} \int_{- \infty}^{\delta}\drm t\,\exp( - t^{2} )$. When we ask for the expected distribution of large-scale structure (an average over small scales), the cosmological principle implies that $\Delta_{i}$ is iid within each averaged patch and we can conclude that the density contrast averaged over large-enough scales is Gau{\ss}ian distributed, which is one of the fundamental insights of modern cosmology.

However, the central limit theorem does not imply convergence to a Gau{\ss} distribution in its tails.  It is also clear that the density contrast has bounded support, being constraint to assume values larger than $-1$. While the distribution of $\delta$ is approximated by a Gau{\ss} distribution in the vicinity of its mean, its tails be very different. They must not necessarily allow for an expansion based on non-Gau\ss ianity parameters [$f^{}_{\rm NL}$, $g^{}_{\rm NL}$ etc.].
\vs{1.5mm}

{\it Extreme-Values: Limiting Distributions\;---\;}Let us now return to the question of the formation of PBHs. For that we are not at all interested in the typical density contrast of a Hubble patch, rather we ask for the probability to find a certain sample maximum{\;---\;}a question about the extrema, thus the most untypical situation. The Fisher-Tippett theorem \cite{fisher_tippett_1928} says that the distribution of {\it sample maxima},
\begin{align}
	M_{N}
		&\equiv
					\underset{i\.=\.1,\.\ldots,\.N}
						{\max\!
						\big(
							\Delta_{i}
						\big)}
					\; ,
					\label{eq:sample-maxima}
\end{align}
necessarily follows generalised extreme-value statistics. These are inherently non-Gau{\ss}ian and have, to our knowledge, so far not been used to describe PBH formation.

Specifically, it can be proven that if there exists sequences $\{ a_{N} > 0 \}_{N\mspace{2mu}=\mspace{2mu}1}^{\infty}$ and $\{ c_{N} \in \Rbb \}_{N\mspace{2mu}=\mspace{2mu}1}^{\infty}$ such that
\begin{align}
	\underset{N\mspace{2mu}\ra\mspace{2mu}\infty}
		{\lim}\.
		{\rm Prob}\!
		\left(
			\frac{ 
				M_{N}
				-
				a_{N} }
			{ c_{N} }
			<
			\delta
		\right)
		&\equiv
					H( \delta )
					\; ,
					\label{eq:Limit-Probability-Fischer-Tippett}
\end{align}
where $H( \delta )$ is a non-degenerate CDF, then this function {\it necessarily} belongs to one of the following classes (\cf~Ref.~\cite{10.5555/262578}):
\begin{align}
	\!H^{s}_{\alpha,\.\gamma}( \delta )
		&=
					\exp{
						\begin{cases}
							-\!
							\left[
								1
								+
								s\!
								\left(
									\frac{ \delta - \alpha }
										{ \gamma }
								\right)
							\right]_{}^{- 1 / s }
								&
									( s \neq 0 )
							\\[4mm]
							-
								\exp{
								\!\left[
									-\!
									\left(
										\frac{ \delta - \alpha }
											{ \gamma }
									\right)
								\right]
								}
								&
									( s = 0 )
						\end{cases}
					}
					\; .
					\label{eq:P-Generalised-EVDs}
\end{align}
The associated probability density function (PDF), $h^{s}_{\alpha,\.\gamma}$, is related to the CDF $H^{s}_{\gamma,\,\alpha}$ via
\vs{-1mm}
\begin{align}
	H^{s}_{\alpha,\.\gamma}( \delta )
		&=
					\int_{- \infty}^{\delta}\d y\;h^{s}_{\alpha,\.\gamma}( y )
					\; .
					\label{eq:PDF-Generalised-EVD-integral}
\end{align}
Above, $s$, $\alpha$ and $\gamma$ are the shape-, location- and scale parameters, respectively. Their values depend on the details of the specific physical situation to be described. If these are not fully known (which is often the case), they have to be inferred from data, \eg, in the fields of finance and climate research, but also in astrophysics, such as in studies of the most massive galactic halos or Galaxy clusters (\cf~Refs.~\cite{2009EL.....8859001A, 2011MNRAS.418L..20H, Chongchitnan:2011eq, Davis:2011vr, Waizmann:2011xu, Reischke:2015jga}). The choices $s = 0$, $s > 0$ and $s < 0$, correspond to the {\it Gumbel}, {\it Fr{\'e}chet}, and {\it Weibull} distributions, respectively.

Any probability distribution of a set of iid random variables, for which Eq.~\eqref{eq:Limit-Probability-Fischer-Tippett} holds, is said to belong to the {\it maximum domain of attraction} of the distribution $H$, of which there are only three subject to the mentioned Fisher-Tippett theorem (see Ref.~\cite{10.5555/262578} (pages 153 -- 157) provides a good respective overview.)
\vs{1.0mm}

\begin{figure}[t]
	\centering
	\vs{3.2mm}
	\includegraphics[width = 0.4208 \textwidth]
		{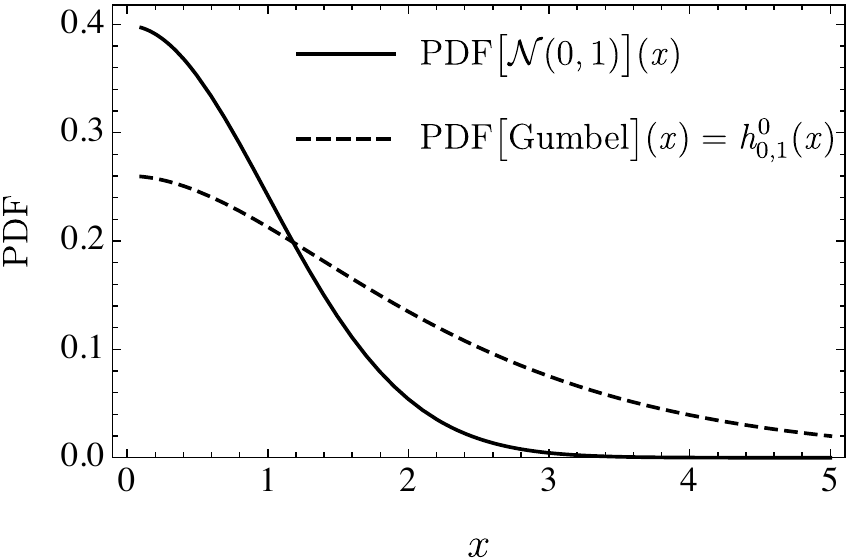}
	\caption{Illustration of the asymptotic behaviour of 
		the Normal and the Gumbel probability 
		distribution functions.}
	\label{fig:Gauss-and-Gauss-Illustration}
\end{figure}

{\it Extreme-Values: The Gau{\ss}ian Case\;---\;}As an illustrative example, let us demonstrate the maximum domain of attraction for the case in which the random variables are Normal distributed, $\Ncal( \mu,\.\sigma )$, \ie
\vs{1mm}
\begin{align}
	{\rm Prob}
	\left(
		\Delta
		<
		\delta_{\crm}
	\right)
		&=
					\Phi\!
					\left( 
						\frac{ \delta_{\crm} - \mu }{ \sigma } 
					\right)
					.
					\label{eq:Prob-Gauss-2}
\end{align}
Then, the choice for the sequences $\{ a_{N} \}_{N\mspace{2mu}=\mspace{2mu}1}^{\infty}$ and $\{ c_{N} \}_{N\mspace{2mu}=\mspace{2mu}1}^{\infty}$ (\cf~Ref.~\cite{Leadbetter1983}),
\begin{subequations}
\begin{align}
	a_{N}
		&=
					\sqrt{2 \ln( N )\,}
					\; ,
					\label{eq:aN-Gauss}
					\\[1.5mm]
	c_{N}
		&=
					a_{N}
					-
					\frac{ \ln\!\big[ \mspace{-1mu}\ln( N ) \big]
						+
						\ln( 4 \pi )}
						{ 2\.a_{N} }
					\; ,
					\label{eq:bN-Gauss}
\end{align}
\end{subequations}
yields
\vs{-0.5mm}
\begin{align}
	H^{0}_{\mu,\.\sigma}( \delta_{\crm} )
		&=
					\exp\!
					\left[
						-
						\exp\!
						\left(
							-\mspace{1mu}
							\frac{
								\delta_{\crm}
								-
								\mu }
							{ \sigma }
						\right)
					\right]
					,
					\label{eq:Limit-Probability-Gauss}
					\\[-4mm]
					\notag
\end{align}
showing that {\it the maxima of Gau{\ss}ian random variables are Gumbel distributed}. We note that, if $\sigma$ is much smaller than $\delta_{\crm} - \mu$, the probability $1 - H^{0}_{\mu,\.\sigma}( \delta_{\crm} )$ develops an exponential tail-behaviour. This behaviour can also be observed for the PDFs, which is illustrated in Fig.~\ref{fig:Gauss-and-Gauss-Illustration} for the Normal and Gumbel distribution. The latter falls off much slower than the former. Note that the mean $\mu$ and variance $\sigma^{2}$ of the Gau{\ss} distribution are \emph{not} the mean $\tilde{\mu}$ and variance $\tilde{\sigma}^{2}$ of the resulting Gumbel distribution, respectively, which are connected via
\vs{-0.5mm}
\begin{align}
	\tilde{\mu}
		&=
					\mu
					+
					\gamma^{}_{\Erm}\.\sigma
					\; ,
	\q\;
	\tilde{\sigma}^{2}
		=
					\zeta( 2 )\,\sigma^{2}
					\; ,
					\label{eq:Average-Variance-Gauss-Gumbel-Relation}
\end{align}
where $\gamma^{}_{\Erm} \approx 0.577$ is the Euler-Mascheroni constant, and $\zeta$ is the Riemann zeta function, with $\zeta( 2 ) = \pi^{2} / 6 \approx 1.64$.

{\it Finite Block Size\;---\;}Having discussed the limiting distributions of extrema, corresponding to infinitely large blocks, we next turn to the corresponding results when the block size is finite. It is really only this case which is applicable in cosmology as the observable Universe does certainly not contain an infinity of Hubble patches.

\begin{figure}[t]
	\centering
	\vs{-2mm}
	\includegraphics[width = 0.42 \textwidth]
		{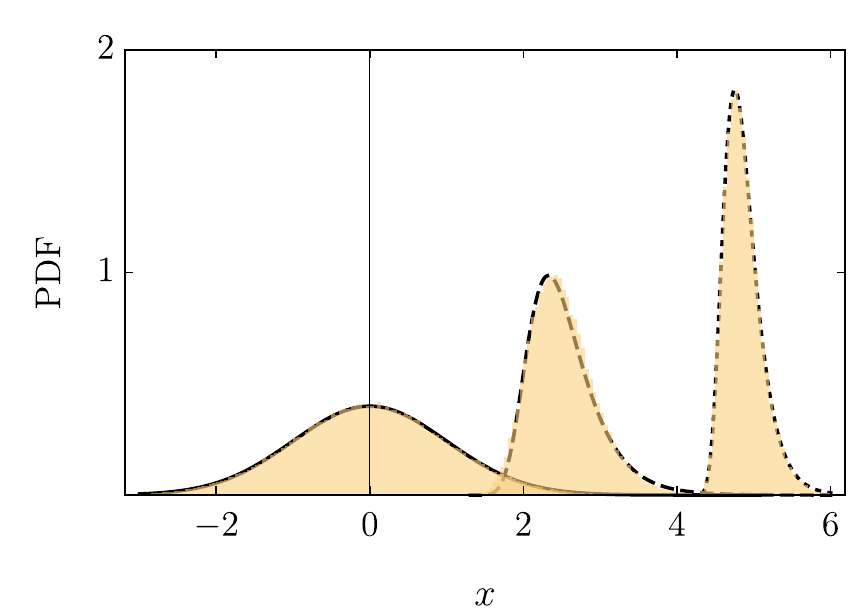}
	\vs{-1.65mm}
	\caption{Probability distribution functions 
			for maxima of $\Ncal(0, 1)$-distributed random variables,
			using blocks of size 
				$N = 1$ (solid curve), 
				$10^{2}$ (dashed curve) and 
				$10^{6}$ (dotted curve),
			with $10^{5}$ realisations each. 
			The latter two cases are well approximated by 
			Eqs.~(\ref{eq:Scaled-Gumbel-Distribution}, \ref{eq:alphaN}{\color{midblue}-c}); 
			the former corresponds to the Gau{\ss}ian case.
			\vs{-2mm}
		}
	\label{fig:Gauss-to-Gumbel-for-finite-N}
\end{figure}

Let us therefore explicitly demonstrate the effect of finite $N$ in a basic illustrative example: The derivation of the statistical distribution of maxima of $\Ncal( \mu,\.\sigma )$-distributed random variables within blocks of size $N$. We should stress that {\it the Normal distribution is recovered for $N = 1$}; for $N > 1$, it is easy to show that the resulting PDF for the maxima is well approximated by the rescaled Gumbel distribution (\cf~Ch.~10.3 of Ref.~\cite{David2004})
\begin{align}
	\tilde{h}_{N}( \delta )
		&\equiv	
					\frac{ 1 }{ \.\gamma_{N} }\.\exp\!
					\left[
						-
						\frac{ \delta - \alpha_{N} }{ \gamma_{N} }
						-
						\exp\!
						\left(
							-
							\frac{ \delta - \alpha_{N} }{ \gamma_{N} }
						\right)
					\right]
					,
					\label{eq:Scaled-Gumbel-Distribution}
\end{align}
with
\vs{-1mm}
\begin{subequations}
\begin{align}
	\alpha_{N}
		&=
					\Phi^{-1}\!
					\left(
						1
						-
						\frac{ 1 }{ N }
					\right)
					\notag
					\displaybreak[1]
					\\[1.5mm]
		&\simeq
					\mu
					+
					\sigma\.
					\sqrt{
						2 \ln( N )
						-
						\ln\!
						\big[
							4 \pi \ln( N )
							-
							2 \pi \ln( 2 \pi )
						\big]\,}
					\; ,
					\vphantom{1_{_{_{_{_{_{1}}}}}}}
					\label{eq:alphaN}
					\displaybreak[1]
					\\[-1.5mm]
	\gamma_{N}
		&=
					\Phi^{-1}\!
					\left(
						1
						-
						\frac{ 1 }{ e\mspace{2mu}N }
					\right)
					-
					\Phi^{-1}\!
					\left(
						1
						-
						\frac{ 1 }{ N }
					\right)
					\notag
					\displaybreak[1]
					\\[1.5mm]
		&\mspace{-20mu}
		\simeq
					\sigma\.
					\sqrt{
						2 \ln( N )
						-
						\ln\!
						\left(
							2 \pi
							\big[
								2 \ln( N )
								+
								2
								-
								\ln( 2 \pi )
							\big]
						\right)
						+ 2 \,}
					\notag
					\displaybreak[1]
					\\[1.5mm]
		&
					-
					\sigma\.
					\sqrt{
						2 \log( N )
						-
						\ln\!
						\big[
							4 \pi \ln( N )
							-
							2 \pi \ln( 2 \pi )
						\big]
					\,}
					\vphantom{1_{_{_{_{_{_{1}}}}}}}
					\; ,
					\label{eq:gammaN}
\end{align}
\end{subequations}
where the asymptotic expansions hold to $\Ocal( 1 / N )$, and $\Phi^{-1}$ is the inverse CDF of the Normal distribution. For any other probability distribution in the maximum domain of attraction of the Gumbel distribution, such as a log-normal one, a corresponding expression holds upon replacing $\Ncal( \mu,\.\sigma )$ \footnote{Here we focus on the Gumbel distribution and its maximum domain of attraction, wherein many of the probability distributions used for PBH studies are contained. While it is not unreasonable to image physical situations which lead to the Fr{\'e}chet distribution (including initial Cauchy, Pareto or Loggamma distributions), it is clear that no physically viable scenario describing PBH formation could lead to the Weibull distribution, as this has bounded support for moderately positive arguments, while being infinitely extended for negative values}. For the Gau{\ss}ian case (corresponding to $N = 1$) as well as for the values $N = 10^{2}$ and $N = 10^{6}$, this is depicted in Fig.~\ref{fig:Gauss-to-Gumbel-for-finite-N}. It can be seen how an increase of $N$ leads to a transition from the Normal to the Gumbel distribution \footnote{In the above example we kept the mean and variance of the distribution of the individual random variables fixed, yielding an $N$-dependence of mean $\tilde{\mu}$ and variance $\tilde{\sigma}^{2}$ of the resulting block-maxima distribution. In practice, however, mostly the opposite case is the relevant one, \ie~fixing the block size to some value $\tilde{N}$ and require $\tilde{\mu}$ and $\tilde{\sigma}^{2}$ to be size-independent. In order to achieve this, the original parameters $\mu$ and $\sigma^{2}$ must assume the values [\cf~Eqs.~(\ref{eq:alphaN},{\color{midblue}b})] $\mu = \tilde{\mu} - \gamma^{}_{\Erm}\.\tilde{\sigma} / \gamma_{\tilde{N}}\.\zeta( 2 ) - \alpha_{\tilde{N}}$ and $\sigma^{2} = \gamma_{\tilde{N}}^{2}\,\tilde{\sigma}^{2} / \zeta( 2 )$, respectively.}.

{\it Mass Distribution\;---\;}Before applying the above machinery to the case of the formation of PBHs, let us briefly comment on the derivation of their mass function. For the case of the Normal distribution, it has been shown in Ref.~\cite{Niemeyer:1997mt} that the fraction $\phi_{\rm PBH}$ of PBHs per logarithmic mass interval is given by
\begin{align}
	\frac{ \d \phi_{\rm PBH}( M ) }{ \d \ln M }
		&\approx
					\frac{ {\rm PDF}\!\left[ \Ncal( \mu, \sigma ) \right]\!
					\big[ \delta( M ) \big] }
					{ 1 - {\rm CDF}\!\left[ \Ncal( \mu, \sigma ) \right]\!
					( \delta_{\crm} ) }\.
					\frac{ \d\delta( M ) }{ \d \ln M }
					\; ,
					\label{eq:initialMF}
\end{align}
being normalised such that
\begin{align}
	\int_{\delta_{\crm}}^{\infty} \d \ln\mspace{-1mu}M\;
	\phi_{\rm PBH}( M )
		&=
					1
					\; .
					\label{eq:nPBH-Normalisation}
\end{align}
Equation \eqref{eq:initialMF} can easily be generalised to other probability distributions.

In order to calculate the PBH mass function, the mass dependence of the density contrast $\delta$ has to be known. Therefore, we remind ourselves that black-hole formation is associated with critical phenomena \cite{Choptuik:1992jv}{\;---\;}the application of which to PBH formation has been studied by various authors (\cf~Refs.~\cite{Koike:1995jm, Niemeyer:1997mt, Evans:1994pj, Kuhnel:2015vtw}). The conclusion is that the mass function has an upper cut-off at around the horizon mass but there is also a low-mass tail \cite{Yokoyama:1998qw}. As demonstrated in Ref.~\cite{Kuhnel:2015vtw} for a variety of inflationary scenarios, critical collapse can have a significant impact on the mass function.

If we assume for simplicity that the density fluctuations have a monochromatic power spectrum and identify the amplitude of the density fluctuation when that scale crosses the horizon, $\delta$, as the control parameter, then the black-hole mass is \cite{Choptuik:1992jv}
\vs{-1mm}
\begin{align}
	M( \delta )
		&=
					k\.M_{\rm H}\.
					\big(
						\delta
						-
						\delta_{\crm}
					\big)^{\tau}
					\qq
					( \delta > \delta_{\crm} )
					\; ,
					\label{eq:M-Critical-Scaling}
					\\[-7.2mm]
					\notag
\end{align}
and being equal to zero otherwise. Here, $M_{\rm H}$ is the horizon mass and the exponent $\tau$ has a universal value for a given equation of state. For a radiation fluid and assuming sphericity one has $\tau \approx 0.36$ as well as $k \approx 4$ \cite{Musco:2008hv, Harada:2013epa}. The value of $\delta_{\crm}$ is sensitive to the shape of the perturbation \cite{Kuhnel:2016exn, Escriva:2019nsa} as well as to non-Gau{\ss}ianity \cite{Atal:2018neu, Kehagias:2019eil}.

{\it Naturally Emerging Blocks\;---\;}As mentioned earlier, the formation of PBHs through collapse of overdensities re-entering the horizon has an extended duration, taking usually a number of e-folds. The concrete time depends on the characteristics of the overdensity as well as the collapsing medium; using Mexican-hat shape and assuming a purely radiation-dominated environment, Ref.~\cite{Musco:2012au} finds a collapse duration of $\Delta N \sim \Ocal( 10 )$ e-folds. This naturally yields the constitution of blocks consisting of a possibly large number Hubble patches which originate from the time of horizon re-entry. Of course, the precise amount of those patches depends on several factors such as the characteristics of the collapsed medium, the profile of the density perturbation as well as their statistics. 

\begin{figure}[t]
	\centering
	\vs{-2mm}
	\includegraphics[width = 0.45 \textwidth]{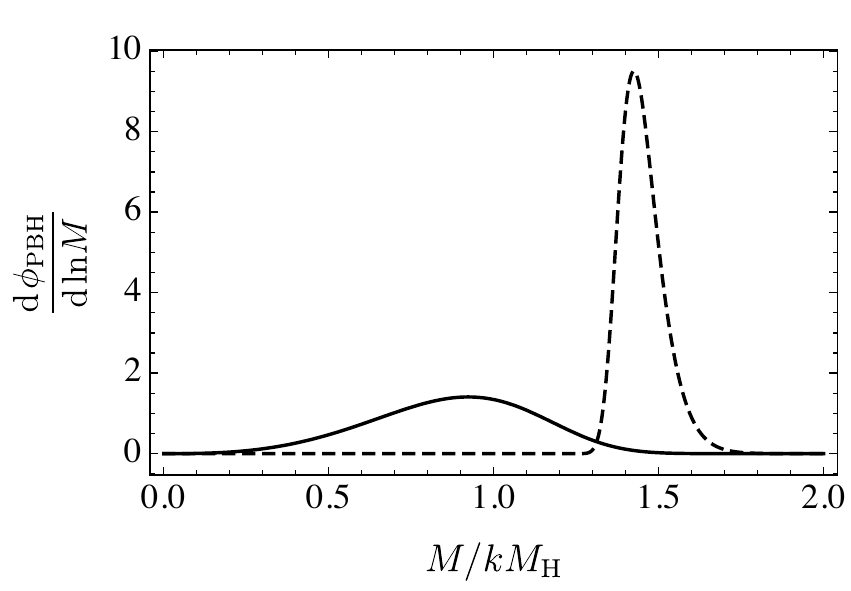}
	\caption{Initial mass function from Eq.~\eqref{eq:initialMF} 
			as a function of PBH mass $M$ in units of $k\.M_{\Hrm}$, 
			with $k \approx 4$ and $\delta_{\crm} = 0.45$, 
			and $M_{\Hrm}$ is the horizon mass 
			at the time of Hubble crossing. 
			The solid curve shows the Gau{\ss}ian case 
			in the instantaneous-collapse approximation
			in comparison to Gumbel case (dashed curve) 
			which takes the finite duration of the collapse into account.
			Both cases utilise the same model of Ref.~\cite{Musco:2012au} 
			as discussed in the main text, 
			yielding an approximate block size of $N_{*} = 10^{3}$.
			}
	\label{fig:Mass-Distribution}
\end{figure}

As an illustrative example, we follow Ref.~\cite{Musco:2012au} and use Gau{\ss}ian statistics for the original overdensities. In turn, the corresponding blocks at PBH formation time respectively contain $N_{*} = ( \Delta N )^{3} \sim 1000$ Hubble patches, each of which initially being subject to a Normal-distributed random variable. It is important to note that any of those blocks will only contain one black hole, where the maximum density contrast triggers its formation. Hence, it is Eq.~\eqref{eq:AtleastonePBH} which gives the probability of finding a black hole within each block; Eq.~\eqref{eq:beta=betatilde/N} yields the probability of having an overdensity collapsing within an original horizon patch. In the instantaneous-collapse approximation both quantities coincide, but that is strictly incorrect. The PBH mass spectrum acquires a $1 / N_{*}$ enhancement and gets significantly modified. This is demonstrated in Fig.~\ref{fig:Mass-Distribution}, which shows that the finiteness of the collapse leads to a peak of the initial mass function Eq.~\eqref{eq:initialMF} at significantly larger mass.

Hence, even for moderate block size one encounters drastic changes to the PBH mass function. In the above example, the underlying statistical distribution of the density contrast was known exactly, which however is often not the case, but that assumption is made nevertheless. However, the results of this work motivate a different {\it best guess}, based on the inevitable emergence of extreme-value distributions whenever block maxima are concerned. This suggests a Gumbel approach to model PBH formation, where a shift towards larger masses as compared to the Gau{\ss}ian case can be expected.

Besides changing the {\it shape} of the mass distribution, in all realistic situations with finite collapse time, \ie~with $N_{*} > 1$, also the {\it abundance} of PBHs will get modified. In order to demonstrate this, we derive the value of the PBH mass density divided by the cosmic background density at the time of PBH formation (\cf~Ref.~\cite{Niemeyer:1997mt})
\vs{-1mm}
\begin{align}
	\Omega_{\rm PBH}\Big|_{\text{form}}
	\mspace{-4mu}
		&=
					\frac{ k }{ N_{*} }\.
					\int_{\delta_{\crm}}^{\infty}\d \delta\;
					( \delta - \delta_{\crm} )^{\tau}\.{\rm PDF}( \delta )
					\; .
					\label{eq:Omega-PBH}
\end{align}
The concrete value for $\Omega_{\rm PBH}$ can readily be derived using a specific probability distribution function ${\rm PDF}( \delta )$.

Let us focus on the Gau{\ss}ian case, discussed in this Section, and set $\mu = 0$. For comparison, we first utilise the so-called horizon-mass approximation, \ie~ignore critical scaling [\cf~Eq.~\eqref{eq:M-Critical-Scaling}], and find
\begin{subequations}
\begin{align}
	\mspace{-5mu}\Omega^{\text{hor mass}}_{\rm PBH}\Big|_{\text{form}}
		&\approx
					0.89\;\erm^{- 0.10 / \sigma^{2}}\.
					\sigma\.
					\big[\.
						1
						+
						\Ocal\big( \sigma^{2} \big)
					\big]
					\; .
					\label{eq:Omega-PBH--horizon-mass-approximation}
\intertext{Using critical and {\it instantaneous} collapse leads to}
	\mspace{-5mu}\Omega^{\text{instant}}_{\rm PBH}\Big|_{\text{form}}
		&\approx
					4.21\;\erm^{- 0.10 / \sigma^{2}}\.
					\sigma^{1.72}\.
					\big[\.
						1
						+
						\Ocal\big( \sigma^{2} \big)
					\big]
					\; .
					\label{eq:Omega-PBH--critical-collapse}
\intertext{{\it Finite collapse time} yields for $N_{*} = 10^{3}$}
	\mspace{-5mu}\Omega^{\text{fin time}}_{\rm PBH}\Big|_{\text{form}}
		&\simeq
					113.2\;\erm^{- 1.58 / \sigma}\.\sigma^{0.36}
					\; ,
					\label{eq:Omega-PBH--finite-time--N*=10to3}
\end{align}
\end{subequations}
being valid to all orders in $\sigma$, but has been derived using Eqs.~\eqref{eq:Scaled-Gumbel-Distribution}, (\ref{eq:alphaN},{\color{midblue}b}) which hold to leading order in $1 / N_{*}$.

In order to obtain the {\it current} value of $\Omega_{\rm PBH}$, the enhancement of the PBH dark-matter fraction, which arises from the formation of the black holes within the radiation-dominated epoch, needs to be taken into account up to the time of matter-radiation equality; it is approximately linear in the cosmic scale factor, being therefore proportional to $1 / \sqrt{M\,}$. For the calculation of $\Omega_{\rm PBH}|_{\text{today}}$, we integrate over masses ranging from the evaporation threshold $3 \times 10^{-19}\,M_{\odot}$ up to the horizon mass at matter-radiation equality $3 \times 10^{17}\,M_{\odot}$. It will become clear below that our results are not very sensitive on these specific values.

For illustration, we consider two cases: a maximally extended continuous spectrum arising from extrapolating the almost scale-invariant primordial power spectrum as measured by CMB observations down to PBH scales, and a monochromatic spectrum \footnote{As being clear from Eq.~\eqref{eq:M-Critical-Scaling}, it is impossible to obtain a final PBH mass spectrum peaked at a single mass. Therefore the term {\it monochromatic} really refers to the single time at which the overdensities, which lead to PBH formation, re-enter the Hubble horizon.}. For the first case we set $\sigma( M ) = \delta_{\rm rms}( M ) = A\,( M / \Msun )^{(1 - n_{\srm}) / 4}$, with the CMB values \cite{Aghanim:2018eyx} $n_{\srm} = 0.97$ and $A = 4 \times 10^{-5}$. This yields $\Omega^{\text{fin time}}_{\rm PBH}|_{\text{today}} \approx 10^{- 10^{4}}$. On the one hand this is very small; on the other hand, when compared to the corresponding result for instantaneous collapse, we find
\begin{align}
	\frac{ \Omega^{\text{fin time}}_{\rm PBH} }
	{ \Omega^{\text{instant}}_{\rm PBH} }\Bigg|_{\text{today}}
	\mspace{-6mu}
		&\simeq
					10^{10^{7}}
					\; ,
					\label{eq:Omega-PBH-ratio-case-i}
\end{align}
which shows an enormous enhancement of the PBH abundance resulting from the slower fall-off of the Gumbel versus the Gau{\ss} distribution. For the monochromatic case, utilising the same amplitude of the primordial power spectrum as above, leads for all conceivable formation times a similar ratio as in Eq.~\eqref{eq:Omega-PBH-ratio-case-i}. Next, assuming that the realistic case of finite formation time yields $100\%$ of PBH dark matter today, implies
\begin{align}
	\frac{ \Omega^{\text{fin time}}_{\rm PBH} }
	{ \Omega^{\text{instant}}_{\rm PBH} }\Bigg|_{\text{today}}
		\simeq
                    \bigg[
                        0.72 \log\!
                        \bigg(\mspace{-2mu}
                            \log\!
                            \bigg[
                                \frac{ M }{ M_{\star} }
                            \bigg]
                        \bigg)
                        -
                        \log\!
                        \bigg(
                            \frac{ M }{ M_{\star} }
                        \bigg)
                    \bigg]^{1.72}
                    \notag
					\\[1.5mm]
					\mspace{-600mu}
					\times\,
					43\,
					\bigg(
					    \frac{ M }{ M_{\star} }
                    \bigg)^{\!\!1/2}
					e^{
					    0.0102\.
					    \big[
                            0.72 \log\!
                            \big(\mspace{-1mu}
                                \log\!\big[
                                    \frac{ M }{ M_{\star} }
                                \big]
                            \big)
                            -
                            \log\!
                            \big(
                                \frac{ M }{ M_{\star} }
                            \big)
					    \big]^{2}
                    }
					\; ,
					\mspace{45mu}
					\label{eq:Omega-PBH-ratio-case-ii}
\end{align}
with $M_{\star} \approx 4.5 \times 10^{23}\.M_{\odot}$. The right-hand side of Eq.~\eqref{eq:Omega-PBH-ratio-case-ii} assumes the values $3 \times 10^{4}$ for $M_{\Hrm} = 1\,M_{\odot}$ and $6 \times 10^{15}$ for $M_{\Hrm} = 10^{-13}\,M_{\odot} \approx 2 \times 10^{20}\,\grm$.

{\it Conclusions\;---\;}In this work we have elaborated on the statistical distributions describing primordial black-hole formation. We argue that in all realistic cases extreme-value distributions must be taken in to account. On the one hand, it is clear that far from the tails, deviations from the Gau{\ss}ian case may well be expected. On the other hand, the finiteness of the collapse naturally leads to consider block maxima of overdense patches which necessarily leads to consider Gumbel or Fr{\'e}chet distributions. As an example we showed that an initial Gau{\ss}ian random density field approximately leads to a Gumbel distribution, with significantly shallower fall-off behaviour. In all conceivable cases for primordial black-hole formation, we find changes in the mass distribution which becomes peaked at larger masses. Furthermore, we obtain an amplification of their current abundance by many orders of magnitude. This underlines the importance of extreme-value statistics for PBH formation, and hence provides new opportunities to naturally explain the dark matter.
\vs{-0.5mm}

\acknowledgements
{\it Acknowledgements\;---\;}We thank Bernard Carr, Cristiano Germani, Filip Linskog and Ilia Musco for helpful discussions. F.K.~acknowledges support from the Swedish Research Council through contract No.~638-2013-8993 during the initial stage of the project. He thanks The Oskar Klein Centre for Cosmoparticle Physics, Bielefeld University and Delta Institute for Theoretical Physics for hospitality and support. D.J.S. acknowledges financial support by Deutsche Forschungsgemeinschaft (DFG) under grant RTG-1620 'Models of Gravity'.
\vs{-2mm}

\setlength{\bibsep}{2.5pt}
\bibliography{refs}

\end{document}